\documentclass[twocolumn,superscriptaddress,preprintnumbers,amsmath,amssymb,prc]{revtex4-2}

\usepackage{graphicx}
\usepackage{dcolumn}
\usepackage{bm}
\usepackage{indentfirst} 
\usepackage{color}
\usepackage{CJK}
\usepackage{booktabs}
\usepackage{float}
\usepackage[colorlinks,linkcolor=blue,urlcolor=blue,citecolor=blue]{hyperref}

\usepackage[colorlinks,linkcolor=blue,urlcolor=blue,citecolor=blue]{hyperref}

\usepackage[colorlinks,linkcolor=blue,urlcolor=blue,citecolor=blue]{hyperref}

\usepackage{tikz,xcolor,hyperref}

\definecolor{lime}{HTML}{A6CE39}
\DeclareRobustCommand{\orcidicon}{
	\begin{tikzpicture}
	\draw[lime, fill=lime] (0,0) 
	circle [radius=0.16] 
	node[white] {{\fontfamily{qag}\selectfont \tiny ID}};
	\draw[white, fill=white] (-0.0625,0.095) 
	circle [radius=0.007];
	\end{tikzpicture}
	\hspace{-2mm}
}
\foreach \x in {A, ..., Z}{%
	\expandafter\xdef\csname orcid\x\endcsname{\noexpand\href{https://orcid.org/\csname orcidauthor\x\endcsname}{\noexpand\orcidicon}}
}
\foreach \x in {A, ..., Z}{%
	\expandafter\xdef\csname orcid\x\endcsname{\noexpand\href{https://orcid.org/\csname orcidauthor\x\endcsname}{\noexpand\orcidicon}}
}

\begin{document}
\begin{CJK*}{UTF8}{gbsn}

\title{Multifractal Dimension Spectrum Analysis for Nuclear Density Distribution}

\author{Weihu Ma(马维虎)}
    \email[Correspondence email address: ]{maweihu@fudan.edu.cn}
    \affiliation{Institute of Modern Physics, Fudan University, Shanghai 200433, People's Republic of China}
\author{Yu-Gang Ma(马余刚)\orcidB{}}
    \email[Correspondence email address: ]{mayugang@fudan.edu.cn}
    \affiliation{Key Laboratory of Nuclear Physics and Ion-beam Application (MOE), Institute of Modern Physics, Fudan University, Shanghai 200433, China}
    \affiliation{Shanghai Research Center for Theoretical Nuclear Physics, NSFC and Fudan University, Shanghai 200438, China}
\author{Wanbing~He(何万兵)}
    \affiliation{Key Laboratory of Nuclear Physics and Ion-beam Application (MOE), Institute of Modern Physics, Fudan University, Shanghai 200433, China}
\author{Bo~Zhou(周波)}
    \affiliation{Key Laboratory of Nuclear Physics and Ion-beam Application (MOE), Institute of Modern Physics, Fudan University, Shanghai 200433, China}
    \affiliation{Shanghai Research Center for Theoretical Nuclear Physics, NSFC and Fudan University, Shanghai 200438, China}

\date{\today} 

\begin{abstract}
We present an integral density method for calculating the multifractal dimension spectrum for the nucleon distribution in atomic nuclei. This method is then applied to analyze the non-uniformity of the density distribution in several typical types of nuclear matter distributions, including the Woods-Saxon distribution, the halo structure and the tetrahedral $\alpha$ clustering. The subsequent discussion provides a comprehensive and detailed exploration of the results obtained. The multifractal dimension spectrum shows remarkable sensitivity to the density distribution, establishing it as an effective tool for studying the distribution of nucleons in nuclear multibody systems.

\end{abstract}


\maketitle

\section{Introduction}
In recent decades, research on nuclear structure and reaction mechanisms \cite{Hongo,Ayyad,YangZH,Yam,Muk,Guo,YangYH,Wang,Huo,ChenPH,ZhouL,Weeb,Cook,Cao,Colo}, especially concerning weakly bound nuclei, has significantly heightened interest in the study of atomic cluster structures. The distinctive quantum multibody properties of weakly bound nuclei, such as halo and skin formation \cite{Tani,Xu,Bagchi,Fang,XuFR,MaZhang,Fang1,Chen,MaCW,Wei}, molecular orbitals, and cluster structures, have garnered widespread attention \cite{bibitem1,WMa,bibitem2,HeWB,Zhou,Ma_NT}. The phenomenon of nucleon clustering within a nucleus is intriguing, giving rise to an alternative view of the nuclear basic structural properties involving the coexistence of mean-field dynamics and clustering dynamics. Consequently, the distribution of nucleons within atomic nuclei exhibits intricate and diverse characteristics.

The distribution of nucleons within atomic nuclei exhibits complexity, non-uniform, and irregularity. This has been preliminary discussed in previous work \cite{bibitem3}. Nucleons are not uniformly distributed within the atomic nucleus; rather, there are regions of high density and relatively sparse regions \cite{Ozawa,Fang2}. The nucleus does not conform to ideal spherical or geometric models, showing irregularities. Various factors influence this spatial distribution, including nucleon interactions, energy levels, and clustering. As a result, the nucleon distribution is complex, uneven, irregular, and displays a range of scales. This characteristic significantly affects our comprehension of nuclear structure, nucleon interactions, dynamics, and reactions.

Nuclear density distribution describes how the nucleons are distributed within the nucleus, which is directly related to the nuclear force because the distribution of nucleons within the nucleus affects how the nuclear interaction operates \cite{Ozawa,Fang2}. A thorough understanding of the nuclear density distribution is crucial for precise modeling and foreseeing the behavior of nuclei, encompassing nucleon interactions within them. This knowledge forms the bedrock for comprehending nuclear processes and properties. The particular organization of nucleons distinctly impacts the potency and extent of the nuclear force.

Fractal objects were introduced to the scientific community by Mandelbrot in 1967 \cite{bibitem4}, characterized chiefly by self-similarity and scale invariant. Multifractal systems extend the concept of fractal systems, requiring more than a single exponent (fractal dimension) to describe their dynamics. Instead, a continuous spectrum of exponents, known as the singularity spectrum, is essential \cite{bibitem5, bibitem6}. The multifractal spectrum's generalized dimensions $D_{q}$ for real exponent $q$ offer a broader parameter than the fractal dimension for describing geometrical properties \cite{bibitem7}. The multifractal spectrum dimension serves as a powerful analytical tool to extract and quantify structural features of complex systems, encompassing non-uniform distributions and irregular shapes. Its applicability extends beyond traditional fractal geometry, enabling the study of irregularities and multi-scale features in complex structures. The multifractal spectrum dimension can quantify the non-uniform distribution density across various regions within complex systems, providing a quantitative measure of this non-uniformity. This facilitates the study of local geometric features within complex systems.

The purpose of this article is to use multifractal spectral dimensions as a tool to study the non-uniformity and irregularity of nucleon distribution within atomic nuclei. This analysis aims to gain a deeper understanding of the complex structure of atomic nuclei and quantify and characterize varying degrees of irregularity and heterogeneity in the distribution of nucleons.

\section{Analysis of Multifractal Spectrum Dimension to Nuclear Mass Distribution Based on Integral Density Method}

Multifractal analysis has proven to be effective in numerous fields \cite{bibitem8,bibitem9,Rod,Tan,Gel,Per,Ahm,Zhao,MaYG, STAR}. The concept of mass multifractal dimension was introduced in works by Refs. \cite{bibitem10,bibitem11,bibitem12,bibitem13,bibitem14}, utilizing the cumulative mass method initially proposed by Ref.~\cite{bibitem15}. Multifractal generalized dimensions, denoted as $D_{q}$, are computed based on the probability distribution function, calculating its moments of varying orders.

The generalized dimensions $D_{q}$ provide a methodology for quantifying the scaling properties and the degree of non-uniformity in mass distributions across diverse scales. Plotting $D_{q}$ against $q$ to generate the multifractal spectrum gives valuable insights into the inherent complexity of the particle distribution. These dimensions, calculated using the Cumulative Mass Method, are defined based on the mass distribution through a scaling relation \cite{bibitem15}
\begin{equation}
\begin{aligned}
\sum_{i=1}(\frac{M_{i}(R)}{M_{0}})^{q}\sim(\frac{R}{R_{0}})^{(q-1)D_{q}},
\end{aligned}
\end{equation}
where the sum over $i$ is a sum over all balls that contain particles, $M_{i}(R)$ is the mass of particles in the $i$-th ball of radius $R$, and $M_{0}$ is the total mass of the system that need to be examined with radius of $R_{0}$. By considering normalized $p_{i}=\frac{M_{i}(R)}{M_{0}}$ as a probability distribution for a randomly chosen particle inside the $i$-th ball of radius $R$, it can be rewritten the left side and then the scale relation is
\begin{equation}
\begin{aligned}
\sum_{i=1}(\frac{M_{i}(R)}{M_{0}})^{q}&=\frac{\sum_{i=1}(\frac{M_{i}(R)}{M_{0}})^{q-1}\frac{M_{i}(R)}{M_{0}}}{\sum_{i=1}\frac{M_{i}(R)}{M_{0}}}\\
&=\langle(\frac{M_{i}(R)}{M_{0}})^{q-1}\rangle\sim(\frac{R}{R_{0}})^{(q-1)D_{q}},
\end{aligned}
\end{equation}
which will hold when the averaging is made with respect to a uniform probability distribution. If all particles have equal mass, then an average over randomly selected particles is equivalent to an average over all particles, so that we can write
\begin{equation}
\begin{aligned}
\langle(\frac{M_{i}(R)}{M_{0}})^{q-1}\rangle=\frac{1}{N}\sum_{k=1}^{N}(\frac{M_{i}(R)}{M_{0}})^{q-1}\sim(\frac{R}{R_{0}})^{(q-1)D_{q}}.
\end{aligned}
\end{equation}
Taking the different mass of each particle into account, the scaling relation of multifractals can be written \cite{bibitem11}
\begin{equation}
\begin{aligned}
\sum_{k=1}^{N_{G}}(\frac{\hat{M_{k}}}{M_{G}})(\frac{M_{k}(R)}{M_{0}})^{q-1}\sim(\frac{R}{R_{G}})^{(q-1)D_{q}},
\end{aligned}
\end{equation}
where $R_{G}$ is the radius of gyration, $N_{G}$ is the number of particles inside $R_{G}$, $\hat{M_{k}}$ is the mass of the $k$-th such particle, $M_{G}$ is the total mass inside $R_{G}$ and $M_{k}(R)$ is the total mass inside a ball of radius $R$ centred on the centre of the $k$-th particle, $M_{0}$ is the total mass of system. The dimensions $D_{q}$ can be obtained from the straight line slopes in a log-log plot of the scaling relation, Eq.(4), using balls of radius $R$ in the range $r < R < R_{G}$ where $r$ is the size of the largest particle in the ball of radius $R$.

The mass fractal dimension, $D_{m}=D_{2}$, for non-uniformly size particle aggregates is then defined by Eq.(4), with $q = 2$. In practical applications the dimensions $D_{q}$ defined by this equation may not be constant over the full range $r < R < R_{G}$ and so estimates of $D_{q}$ from the slopes of straight line portions in log-log plots are employed. When the mass multifractals are embedded into three dimensions, the condition of $1 \leq D_{m} \leq 3$ must be satisfied.

\begin{figure}[!htb]
\centering
\label{fig1}
\includegraphics[width=0.80\hsize]{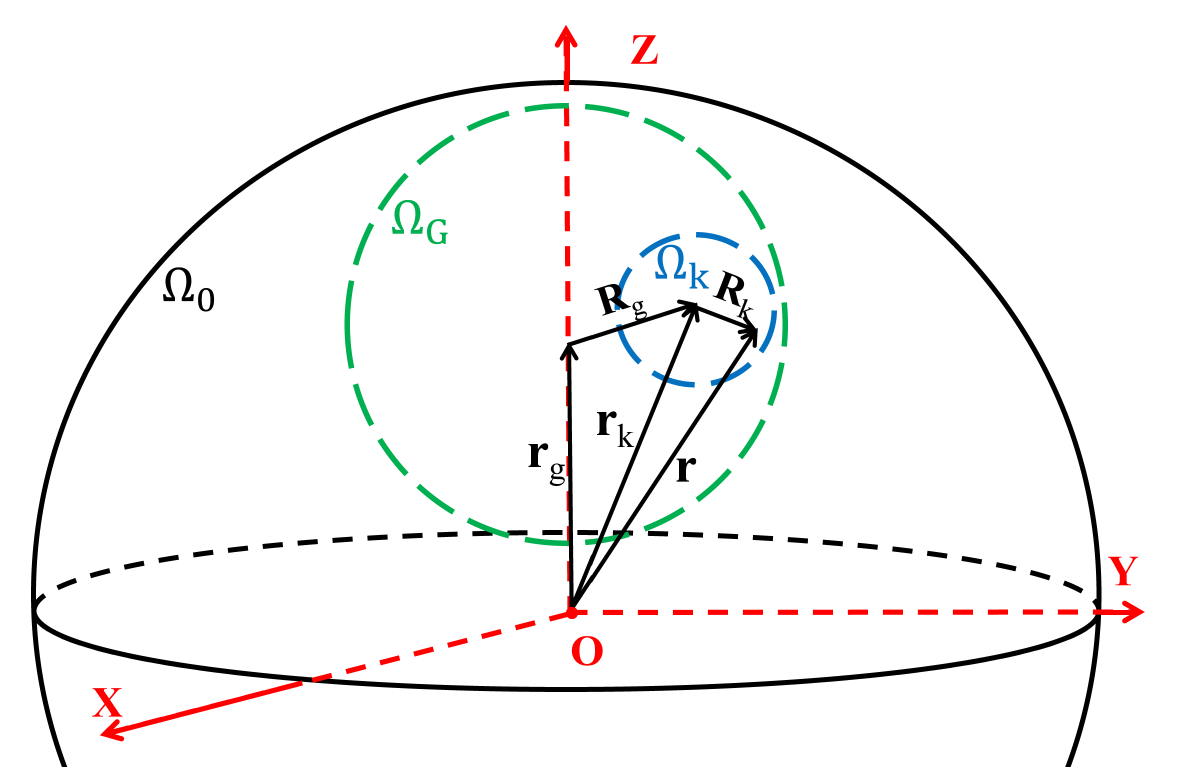}
\caption{(color online) Sketch map for vectors inside a nucleus. Details of captions are explained in the text.}
\label{fig1}
\end{figure}

\begin{figure}[!htb]
\centering
\label{fig2}
\includegraphics[width=0.80\hsize]{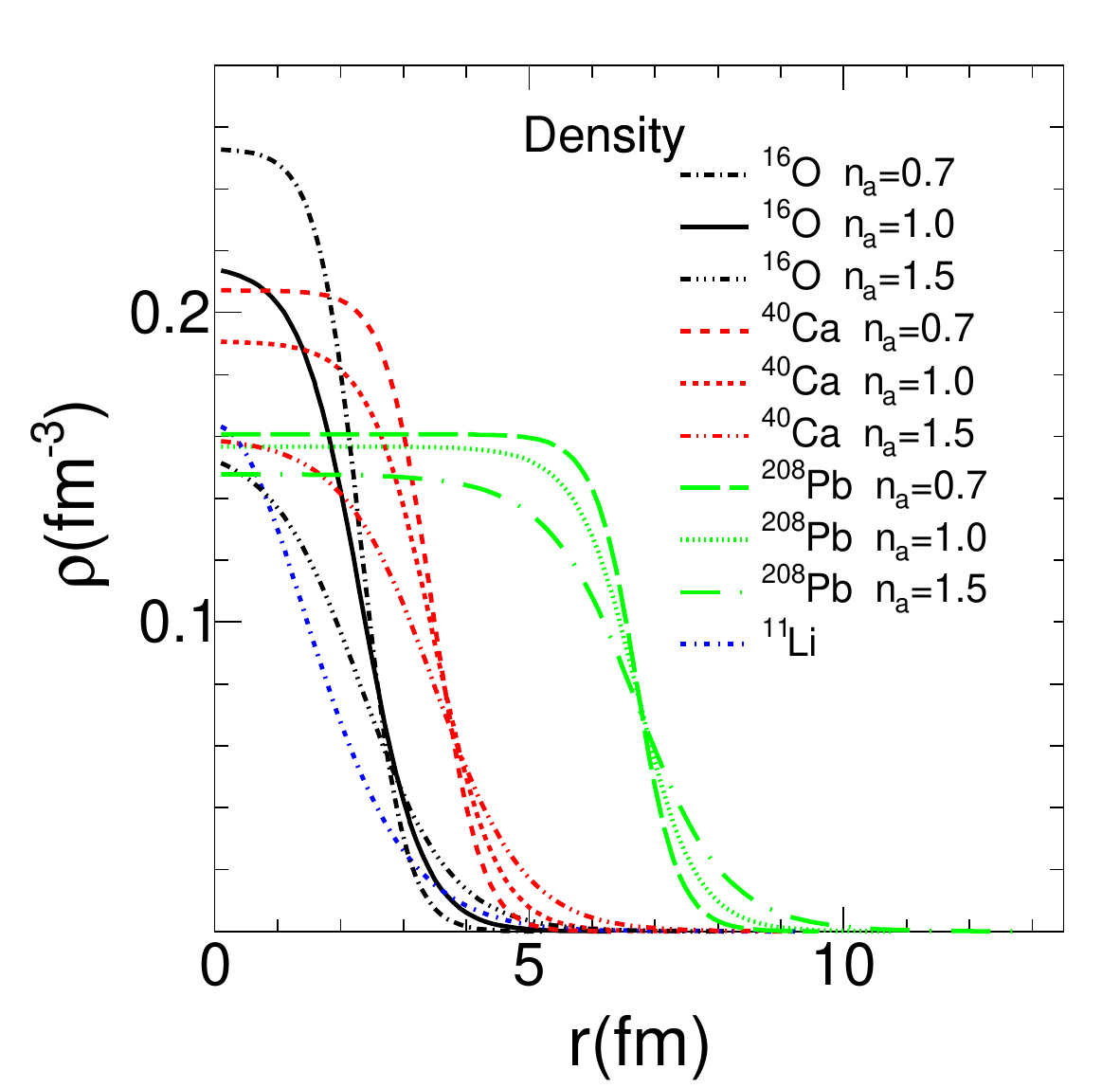}
\caption{(color online) Woods-Saxon type density distributions for nuclei $^{16}$O, $^{40}$Ca, and $^{208}$Pb, and the density distribution for halo nucleus $^{11}$Li given in \cite{bibitem17}. The parameter $n_{a}$ is for scaling the diffuseness parameter $a$.}
\label{fig2}
\end{figure}

To estimate the multifractal dimension of a nuclear system, we utilize the density distributions of protons, denoted as $\rho_{p}(\vec{r})$, and neutrons, denoted as $\rho_{n}(\vec{r})$. The numbers of protons and neutrons, denoted as $Z=\int_{\Omega_{0}}\rho_{p}(\vec{r})d\vec{r}$ and $N=\int_{\Omega_{0}}\rho_{n}(\vec{r})d\vec{r}$ respectively. Here, $\Omega_{0}$ represents the spatial domain of the nuclear system. The density distribution of nucleons is generally obtained by solving the Schr$\ddot{o}$dinger equation for the atomic nuclear system, where the density $\rho(\vec{r})$ is proportional to the sum of squared magnitude of the wavefunction of nucleons $\psi_{i}(\vec{r})$: $\rho(\vec{r})\sim \sum|\psi_{i}(\vec{r})|^{2}$. As shown in FIG. 1, we can get the following representation for calculating Eq. (4).
\begin{equation}
\begin{aligned}
M_{0}=\sum_{\mu=p,n}m_{\mu}\int_{\Omega_{0}}\rho_{\mu}(\vec{r})d^{3}\vec{r},
\end{aligned}
\end{equation}
where the integral is for the whole space; $m_{p}=M_{p}-bi$ and $m_{n}=M_{n}-bi$ with $M_{p}$ and $M_{n}$ being the mass of free proton and neutron respectively and $bi$ being the binding energy per nucleon.

\begin{equation}
\begin{aligned}
M_{G}(\vec{r}_{g})&=\sum_{\mu=p,n}m_{\mu}\int_{\Omega_{G}}\rho_{\mu}(\vec{r})d\Omega_{G}\\
&=\sum_{\mu=p,n}m_{\mu}\int_{\Omega_{G}}\rho_{\mu}(\vec{r}_{g}+\vec{R'}_{g})d^{3}\vec{R'}_{g},
\end{aligned}
\end{equation}
where $\vec{r}=\vec{r}_{g}+\vec{R'}_{g}$ with $\vec{r}_{g}$ is the position vector of centre of ball $\Omega_{G}$ of radius $R _{G}$ and $\vec{R'}_{g}$ is the position vector relative to the centre of ball $\Omega_{G}$.

\begin{equation}
\begin{aligned}
M_{k}^{\mu=p,n}(\vec{r}_{k})=m_{\mu}\int_{\Omega_{k}}\rho_{\mu}(\vec{r}_{k}+\vec{R}_{k})d^{3}\vec{R}_{k},
\end{aligned}
\end{equation}
where $\vec{r}_{k}$ is the position vector of centre of ball $\Omega_{k}$ of radius $R$ and $\vec{R}_{k}$ is the position vector relative to the centre of ball $\Omega_{k}$; $\vec{R}_{g}$ is the position vector of the centre of ball $\Omega_{k}$ relative to the centre of $\Omega_{G}$; $\vec{r}_{k}=\vec{r}_{g}+\vec{R}_{g}$, $|\vec{R}_{g}+\vec{R}_{k}|\leq{R_{G}}$ and $|\vec{R}_{k}|\leq{R}$.

Considering the representations introduced above, then the scaling relation for nuclear system can be written to
\begin{widetext}
\begin{equation}
\begin{aligned}
\frac{\sum_{\mu=p,n}m_{\mu}\int_{\Omega_{G}}(\sum_{\nu=p,n}m_{\nu}\int_{\Omega_{k}}\rho_{\nu}(\vec{r}_{k}+\vec{R}_{k})d\vec{R}_{k})^{q-1}\cdot\rho_{\mu}(\vec{r}_{k})d\vec{r}_{k}}{\sum_{\zeta=p,n}m_{\zeta}\int_{\Omega_{G}}\rho_{\zeta}(\vec{r}_{g}+\vec{R'}_{g})d\Omega_{G}\cdot{M_{0}}^{q-1}}\sim(\frac{R}{R_{G}})^{(q-1)D_{q}};
\end{aligned}
\end{equation}
\end{widetext}
and
\begin{widetext}
\begin{equation}
\begin{aligned}
\frac{\sum_{\mu=p,n}m_{\mu}\int_{\Omega_{G}}\{\ln{(\sum_{\nu=p,n}m_{\nu}\int_{\Omega_{k}}\rho_{\nu}(\vec{r}_{k}+\vec{R}_{k})d\vec{R}_{k})}-\ln{M_{0}}\}\cdot\rho_{\mu}(\vec{r}_{k})d\vec{r}_{k}}{\sum_{\zeta=p,n}m_{\zeta}\int_{\Omega_{G}}\rho_{\zeta}(\vec{r}_{g}+\vec{R'}_{g})d\Omega_{G}}\sim D_{1}\cdot\ln{(\frac{R}{R_{G}})},
\end{aligned}
\end{equation}
\end{widetext}
for $q=1$.
This integral density approach enables the estimation of the multifractal dimension of a nuclear system, offering valuable insights into the non-uniformity nature of the nuclear structure. Employing this methodology, this article delves into exploring the multifractal dimension spectra of various distinct nuclear distribution patterns. These patterns encompass typical structures like Woods-Saxon type nuclei such as $^{16}$O, $^{40}$Ca, and $^{208}$Pb, as well as nuclei with specific configurations like the halo structure in $^{11}$Li and the tetrahedral alpha cluster structure in $^{16}$O.

\begin{figure}[!htb]
\label{fig3}
\includegraphics[width=0.80\hsize]{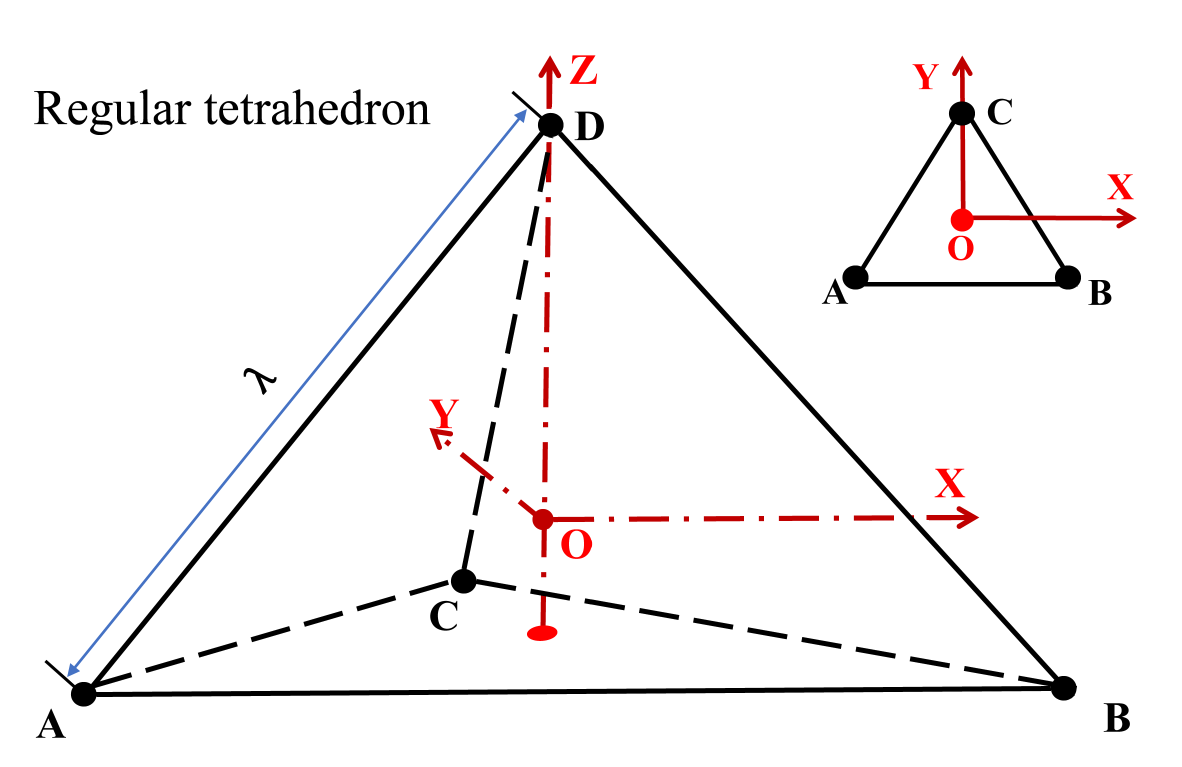}
\caption{(color online) Coordinate representations of the $^{16}$O with tetrahedral cluster structure.}
\label{fig3}
\end{figure}

\begin{figure*}[!htb]
\label{fig4}
\includegraphics[width=0.95\hsize]{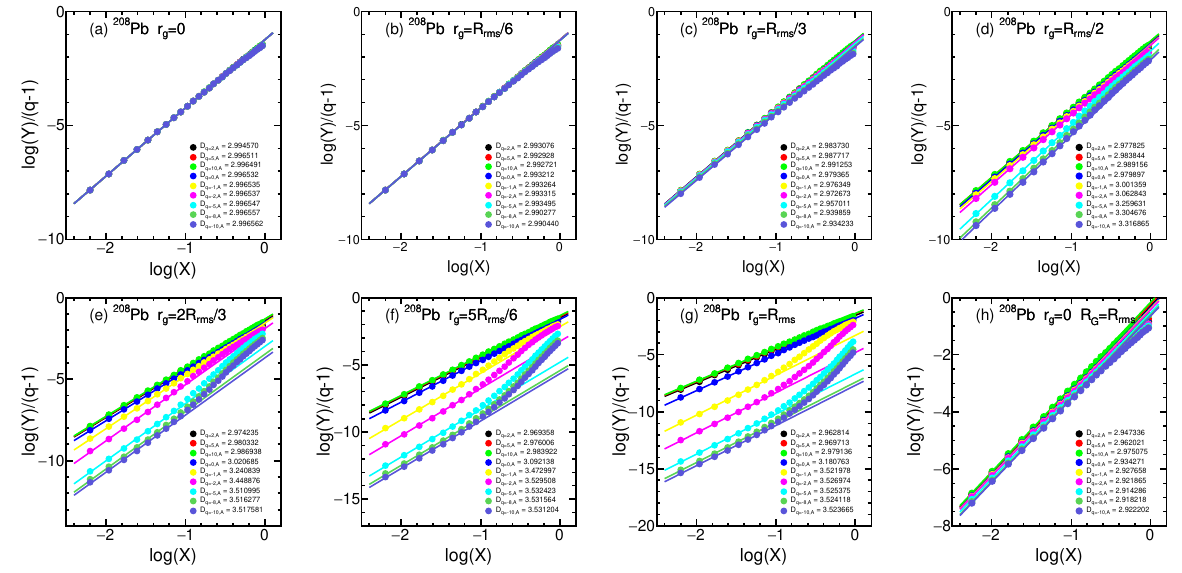}
\caption{(color online) Estimation of $D_{q,A}$ for $q=2, 5, 10, 0, -1, -5, -8, -10$ from the slopes of the plots $log(Y)/(q-1)$ versus $log(X)$, where $Y$ is the left side of equation (8) and $X=R/R_{G}$ with $R_{G}=3(R_{rms}-1.2)/5+1.2$. The subscript $A$ of $D_{q,A}$ means considering protons and neutrons together. Figures (a)$\sim$(g) are respectively corresponding to $r_{g}=0$, $R_{rms}/6$, $R_{rms}/3$, $R_{rms}/2$, $2R_{rms}/3$, $5R_{rms}/6$, and $R_{rms}$ with $R_{rms}=\sqrt{\frac{Z}{A}R_{0p}^{2}+\frac{N}{A}R_{0n}^{2}}$. Figures (h) is corresponding to $r_{g}=0$ and $R_{G}=R_{rms}$.}
\label{fig4}
\end{figure*}

\begin{figure*}[!htb]
\centering
\label{fig5}
\includegraphics[width=0.90\hsize]{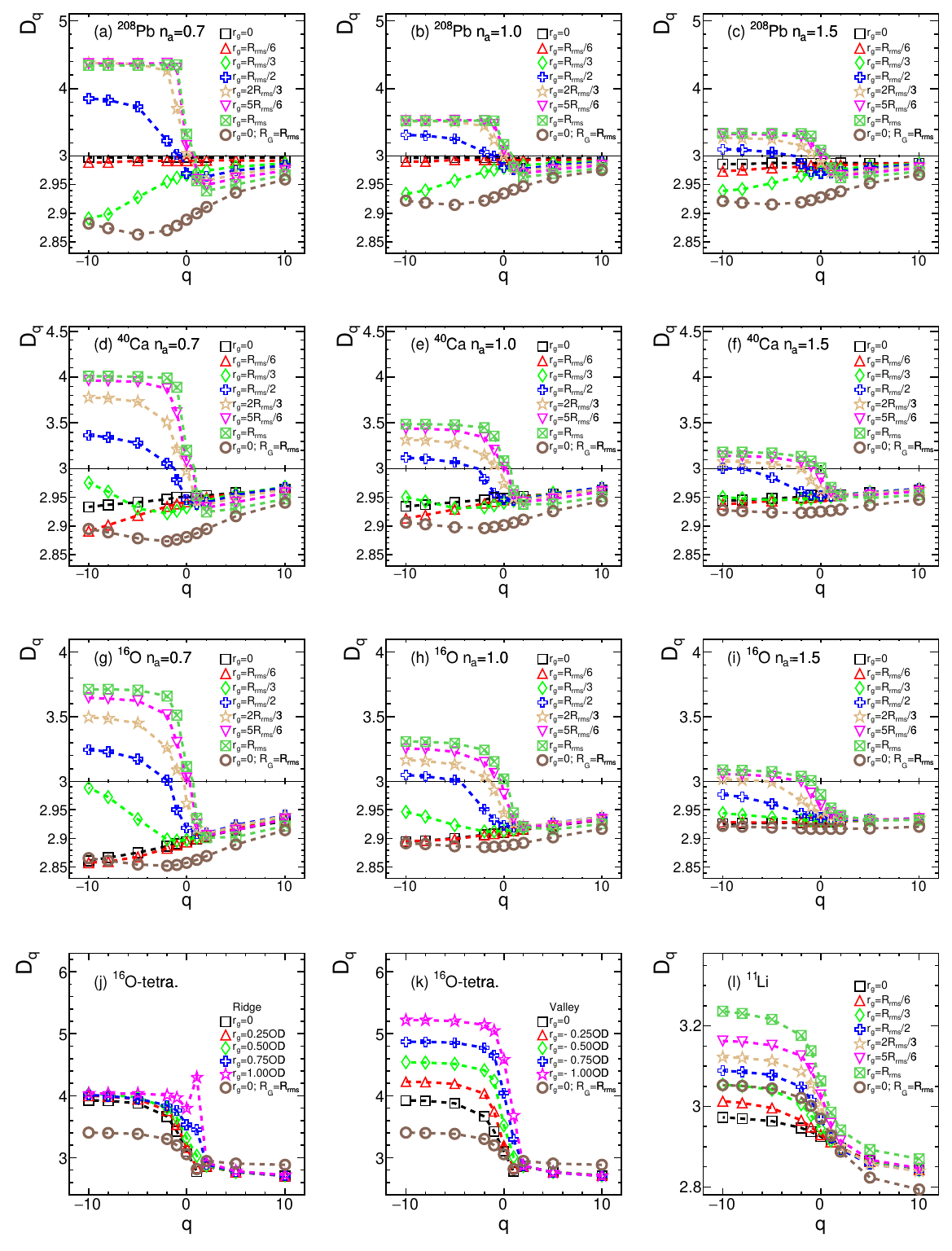}
\caption{(color online) The $D_q$ spectrum, relative to the parameter $q$, is presented in Figures (a) to (c) for $^{208}$Pb with varying diffuseness parameters. Similarly, Figures (d) to (f) illustrate the $D_q$ spectrum for $^{40}$Ca, and Figures (g) to (i) depict the spectrum for $^{16}$O. Break scales are used from Figures (a) to (i), i.e. different scale in Y-axis above or below 3. In Figure (j), the $D_q$ spectrum is displayed for $^{16}$O with a tetrahedral structure along the Ridge direction (OD), while Figure (k) shows the spectrum for $^{16}$O with a tetrahedral structure along the Valley direction (-OD). Finally, Figure (l) demonstrates the $D_q$ spectrum for $^{11}$Li with a halo structure.}
\label{fig5}
\end{figure*}

The nuclei $^{16}$O, $^{40}$Ca, and $^{208}$Pb with density distribution of Woods-Saxon type \cite{bibitem16}
\begin{equation}
\begin{aligned}
\rho(r)=\frac{\rho_{0}}{1+e^{(r-R_{0})/a}}
\end{aligned}
\end{equation}
are taken into account as examples to examine the scaling relation and extract their multifractal spectrum dimensions. The normalization coefficient $\rho_{0}$, diffuseness parameter $a$, and radius $R_{0}$ are satisfying normalization condition
\begin{equation}
\begin{aligned}
4\pi\int_{0}^{\infty}\rho(r)r^{2}dr=W
\end{aligned}
\end{equation}
where $W$ could be the number of protons $Z$, neutrons $N$, or nucleons $A=N+Z$. The $R_{0}$ parameters for proton and neutron distributions are given to be $R_{0p}$ and $R_{0n}$ respectively, where $R_{0p}=1.81Z^{1/3}-1.12$; $R_{0n}=1.49N^{1/3}-0.79$. The $a$ parameter for proton and neutron distributions are given to be $a_{p}=0.47-0.00083Z$ and $a_{n}=0.47+0.00046N$ respectively. The diffuseness parameter $a$ of the nuclei $^{16}$O, $^{40}$Ca, and $^{208}$Pb are scaled by parameter $n_{a}$ for comparison. The plots for Woods-Saxon type density distributions of nuclei $^{16}$O, $^{40}$Ca, and $^{208}$Pb, and the density distribution for halo nucleus $^{11}$Li given in \cite{bibitem17} are shown in FIG. 2.

It has been pointed out that atomic nuclei with tetrahedral symmetry could be encountered all over the nuclear chart \cite{bibitem18}. Tetrahedral structure $\alpha$+$\alpha$+$\alpha$+$\alpha$ of $^{16}$O was suggested in Ref. \cite{bibitem19,HeWB,LiYA,Ding,Ma_NT,Ma_SCP}. The rotation-vibration spectrum of a 4$\alpha$ configuration with tetrahedral symmetry $T_{d}$ and the evidence for the occurrence of this symmetry in the low-lying spectrum of $^{16}$O were discussed and verified by experimental excitation spectrum. It is possible that atomic nuclei with tetrahedral symmetry have a definite interest in all related fields of physics.

In order to exam the scale behavior of $^{16}$O with tetrahedral cluster structure shown in FIG. 3, with the distribution of nucleons within an alpha particle based on the alpha-particle model is Gaussian type \cite{bibitem20}
\begin{equation}
\begin{aligned}
\rho(\vec{r})&=\sum_{i=1}^{4}m_{i}(\frac{\delta_{i}}{\pi})^{3/2}e^{-\delta_{i}(\vec{r}-\vec{r}_{i})^{2}},
\end{aligned}
\end{equation}
where, $m_{i}=m_{\alpha}$; $\delta_{i}=\delta_{\alpha}=0.52$; $\vec{r}_{i}$ is the vector of $\alpha$-cluster relative to the centre of mass of the $^{16}$O. In FIG. 3, the center points of each $\alpha$ cluster respectively are $A=\{\beta,\frac{\pi}{2}+arctan(\frac{\sqrt{2}}{4}),\frac{7\pi}{6}\}$; $B=\{\beta,\frac{\pi}{2}+arctan(\frac{\sqrt{2}}{4}),\frac{11\pi}{6}\}$; $C=\{\beta,\frac{\pi}{2}+arctan(\frac{\sqrt{2}}{4}),\frac{\pi}{2}\}$; $D=\{\beta,0,0\}$; where $\beta=\frac{\sqrt{6}}{4}\lambda$ and $\lambda$ is the edge-length (the center distance of arbitrary two $\alpha$ clusters). $\beta=r_{i}$ for tetrahedron structure of $^{16}$O.

\begin{figure*}[!htb]
\centering
\label{fig6}
\includegraphics[width=0.90\hsize]{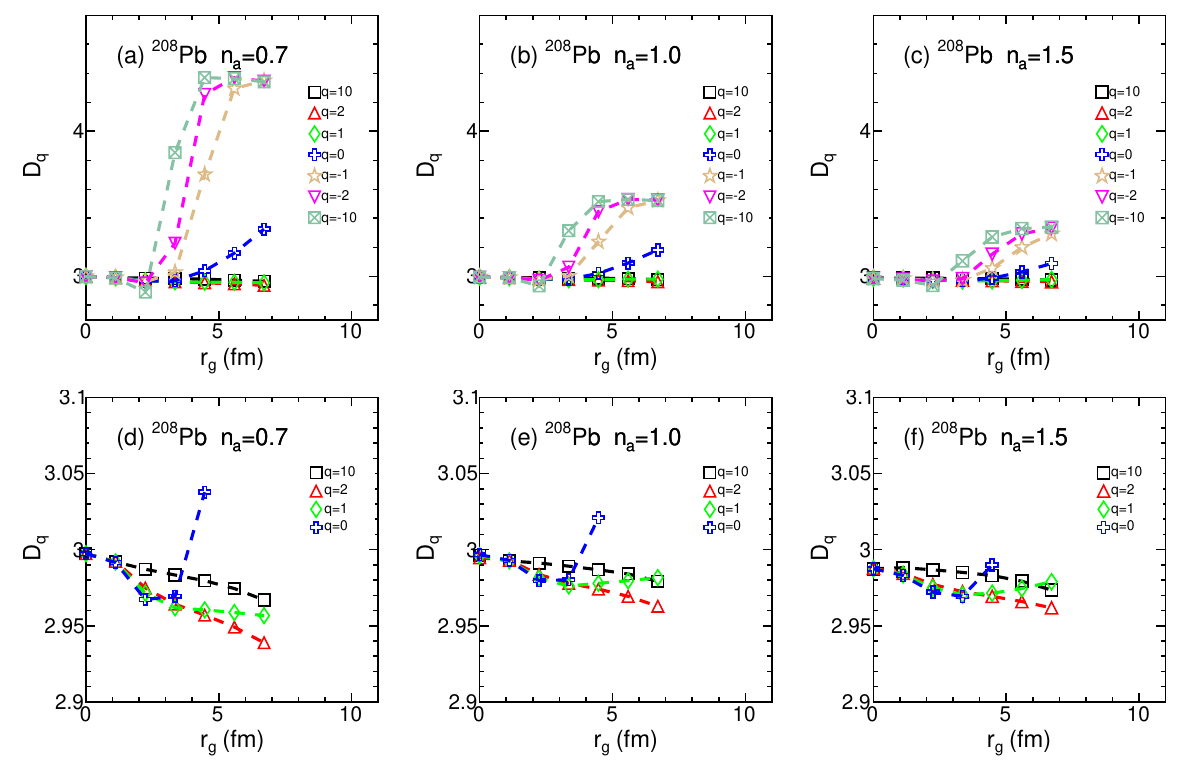}
\caption{(color online) (a)$\sim$(c): the $D_q$ spectrum for nucleus $^{208}$Pb are plotted with respect to the radius $r$ for each specified value of $q$. (d)$\sim$(f): zoom scale for clear view of the $D_q$ spectrum for $q\geq0$.}
\label{fig6}
\end{figure*}

We select seven balls with radius given by $R_{G}=\frac{3(R_{rms}-1.2)}{5}+1.2$. Their centers are placed at radial coordinates $r_{g}=0, \frac{R_{rms}}{6}, \frac{R_{rms}}{3}, \frac{R_{rms}}{2}, \frac{2R_{rms}}{3}, \frac{5R_{rms}}{6},$ and $R_{rms}$. Here, $R_{rms}=\sqrt{\frac{Z}{A}R_{0p}^{2}+\frac{N}{A}R_{0n}^{2}}$ is the square root mater radius. Additionally, we consider one ball with a radius of $R_{G}=R_{rms}$ and position its center at radial coordinate $r_{g}=0$. This setup allows us to extract multifractal dimensions and analyze the scaling relation. The multifractal dimensions $D_{q}$ are determined from the slopes of the linear segments observed in log-log plots. Using $^{208}$Pb as an example, as illustrated in FIG. \ref{fig4} (a)-(g), we observe that the estimated values of $D_{q}$ are nearly identical at $r_{g}=0$, and the discrepancies among them increase with $r_{g}$. This is due to the Woods-Saxon type density distribution (as shown in FIG. 2), demonstrating approximate uniformity at small radii and dispersion at larger radii. FIG. \ref{fig4} (h) represents an overall measurement of $D_{q}$ that includes the majority of the mass distribution. In the case of a deterministic fractal object possessing a unique fractal dimension, like the Sierpinski gasket, the estimated $D_{q}$ values with respect to $q$ are indistinguishable. For the Woods-Saxon type density distribution of $^{208}$Pb, it exhibits a monofractal scaling relation around its center and transitions to a multifractal scaling relation at the edge region.

The $D_{q}$ vs $q$ spectra provide valuable insights into scaling patterns. Typically, the spectrum exhibits a decreasing trend, particularly sigmoidal around $q=0$. This spectrum's variation aids in distinguishing different patterns. Usually, the $D_{q}$ spectra are derived from multifractal analysis depicting non-fractals, mono-fractals, and multi-fractals. Notably, non-fractals and mono-fractals tend to have flatter $D_{q}$ spectra compared to multifractals. For non-fractals, the mono dimension closely aligns with the geometric dimension of integers. The $D_{q}$ vs $q$ graph's distinct patterns are crucial for analyzing and differentiating non-fractals, mono-fractals, and multifractals, highlighting their respective dimensional characteristics.

In FIG. 5 (a)-(i), multifractal dimensions spectrum $D_q$ of $^{16}$O, $^{40}$Ca, and $^{208}$Pb with Woods-Saxon type density distribution are strongly depending on the values of density and the degrees of non-uniformity in density distribution. $D_q$ are discussed in detail in different areas and with different scales of diffusion parameters.

In the central region ($r_g=0$), the density distribution of $^{208}$Pb closely resembles a uniform distribution, resulting in a flat multifractal dimensions spectrum $D_{q}$ at around 2.996, like a liquid droplet phase with $D_{q}\sim3$. A smaller diffusion parameter scale exhibits more uniformity and higher density, leading to a larger value of $D_{q}$ (approximately 2.997) with a flat trend. Conversely, a larger diffusion parameter scale shows less uniformity and lower density, resulting in a smaller value of $D_{q}$ (around 2.987), also with a flat trend. In comparison to $^{208}$Pb, $^{40}$Ca and $^{16}$O have progressively higher densities but less uniform density distribution in the central area. Consequently, their $D_{q}$ values successively decrease and do not remain flat. The change in $q$ values from large to small (from $q>0$ to $q<0$ direction) is progressively suppressed. In general,
in regions with a uniform and higher-density distribution, the $D_{q}$ value is larger and remains independent of $q$ (like $^{208}$Pb at $r_g=0$ with a smaller scale of $a$).
In regions with a uniform and lower-density distribution, the $D_{q}$ value is smaller and remains independent of $q$ (like $^{208}$Pb at $r_g=0$ with a larger scale of $a$).
In regions with a more non-uniform and higher-density distribution, the $D_{q}$ value is smaller and decreases fast as $q$ becomes smaller (like $^{16}$O and $^{40}$Ca at $r_g=0$ with a smaller scale of $a$).
In regions with a less non-uniform and lower-density distribution, the $D_{q}$ value is larger and decreases slowly as $q$ becomes smaller (like $^{16}$O and $^{40}$Ca at $r_g=0$ with a larger scale of $a$).

There is a mechanism for a combined effect of the density magnitude and the degrees of non-uniformity in the density distribution. This mechanism is clearly reflected in the region where $r_{g} = R_{rms}/3$. The case of the area at $r_{g} = 0$ with $R_{g} = R_{rms}$ also demonstrates a similar mechanism, where it encompasses a sizable area exhibiting a variation in the non-uniformity of density distribution. Compared to the former, the latter has a lower average density, leading to a more depressed $D_{q}$ spectrum. The latter also exhibits a similar variation in the tendency of $D_{q}$ spectrum. In the distribution of $^{208}$Pb, the latter contains more non-uniform content, making the curve of the $D_q$ spectrum have a clear turning point. $D_{q}$ decreases with a decrease in $q$ and increases after the turning point with a decrease in $q$. When comparing to the distribution of $^{208}$Pb, for $^{16}$O and $^{40}$Ca, as $q$ decreases, the turning point of the $D_{q}$ curve appears earlier, and the changing trend of the $D_{q}$ curve is steeper. With a larger diffusion parameter scale, the distribution tends to become more uniform, and the change rate in the density distribution becomes more gradual. This results in gentler changes in the $D_{q}$ spectrum curve as well.

For a given atomic nucleus with a Woods-Saxon type density distribution, as the examined density area changes from the central area to the edge area, the degree of non-uniformity rapidly increases, and the density decreases. When examining the area at the position where $r_{g}=R_{rms}/2$, $2R_{rms}/3$, $5R_{rms}/6$, and $R_{rms}$, the $D_{q}$ values from $q=0$ to lower are rapidly amplified and then trend to flat. In the examined region, a larger portion experiences a higher rate of density variation and lower density magnitude, resulting in a more pronounced amplification of $D_{q}$ values for $q\leq0$.

In FIG. 5 (l), the multifractal dimension spectrum $D_{q}$ of $^{11}$Li with a halo structure, is predominantly influenced by a combination of low density and non-uniform density distribution. The $D_{q}$ spectrum, representing various scenarios, demonstrates typical characteristics of a multifractal spectrum. Within the central region ($r_{g}=0$), the behavior of the $D_{q}$ spectrum varies for two balls of different sizes characterized by radii $R_{G}$. The larger ball, associated with a lower average density, yields lower $D_{q}$ values for $q>1$ in comparison to the smaller-sized ball. The difference stems from the larger ball encompassing a more substantial portion of non-uniformity and lower magnitude, thereby significantly amplifying the $D_{q}$ values for $q<1$. When examining the area changing from the central area to the edge area, $D_{q}$ spectra are affected by the combination of density magnitude and the degrees of non-uniformity.

In FIG. 5 (j)-(k), the multifractal dimension spectrum $D_{q}$ of $^{16}$O with a tetrahedral structure, denoted as $\alpha$+$\alpha$+$\alpha$+$\alpha$, exhibits a clear distinction between the analysis conducted along the ridge (OD) and the valley (-OD) directions originating from the center. This disparity is strongly influenced by the density distribution inherent to the tetrahedral $\alpha$ cluster structure. Within the central region, the density distribution is shaped by the combined contributions of Gaussian tails from four $\alpha$ clusters centered at positions A, B, C, and D, as illustrated in FIG. 3. Notably, this distribution is non-uniform in nature. When analyzing along the ridge direction, the trajectory traverses across the peaks of the Gaussian distribution associated with the $\alpha$ clusters, resulting in an abnormal increase of $D_{q}$ for $q=1$ near the position of the peak, indicating significant largeness and non-uniformity in the density distribution. Conversely, in the valley direction, the trajectory follows the superposition of Gaussian tails from three $\alpha$ clusters centered at positions A, B, and C. In this case, the $D_{q}$ for $q\leq1$ are successively significantly magnified from center to edge ($r_{g}=0$, -0.25OD, -0.5OD, -0.75OD, and -OD) due to the low density and non-uniform distribution. This amplification is notably more pronounced than that observed along the ridge direction. These investigations were conducted using a ball with a radius of $R_{G}=3[(\beta+r_{\alpha})-1.2]/5+1.2$. In the central area ($r_{g} = 0$), when compared to a larger ball with $R_{G}=\beta+r_{\alpha}$, both scenarios exhibit an abnormal decrease in $D_{q}$ for $q=1$, indicating significant non-uniformity in the density distribution. Here, $r_{\alpha}$ represents the radius of the $\alpha$ cluster. This nuanced examination provides valuable insights into the multifractal dimension spectrum with respect to the tetrahedral $\alpha$+$\alpha$+$\alpha$+$\alpha$ structure, emphasizing the impact of density distribution intricacies on the observed differences between the ridge and valley directions.

The $D_{q}$ spectrum for nucleus $^{208}$Pb are plotted in FIG. 6 with respect to the radius $r$ for some specified value of $q$. As the density distribution changes from the center to the edge along the radius $r$, distinct trends in the change of $D_{q}$ for different $q$ are apparent, each reflecting the characteristics of the density distribution to varying degrees. For $q\geq2$, the $D_{q}$ values exhibit a monotonically decreasing pattern as the radius $r$ increases, and the larger the value of $q$, the slower the rate of decrease. Upon comparing different diffusion parameters, it becomes evident that larger diffusion parameters, in comparison to smaller ones, correspond to a reduction in density within the square root of the mater radius and are relatively gradual in density decreasing with $r$ increasing, resulting in a relatively sluggish decline in $D_{q}$ as the radius expands. The variation curve of $D_{q}$ for $q\leq1$ demonstrates a pronounced sensitivity to density non-uniformity, with this sensitivity becoming more prominent as $q$ decreases. Initially, the $D_{q}$ values decrease, followed by a rapid increase as the density non-uniformity in the inspected area intensifies. Furthermore, the greater the non-uniformity in density within the inspected area, the higher the observed sensitivity of the $D_{q}$ values.

\section{Summary}
In conclusion, the multifractal spectrum dimension proves to be a valuable tool for understanding the complex structural characteristics inherent in various systems. Its broad applicability provides a precise means to quantify and analyse non-uniform distribution patterns and irregular shapes within atomic nuclei. Using the multifractal dimension spectrum $D_q$, we have carried out a comprehensive analysis of $^{16}$O, $^{40}$Ca and $^{208}$Pb nuclei characterised by the Woods-Saxon type density distribution, as well as the halo structure of $^{11}$Li and the tetrahedral $\alpha$ clustering of $^{16}$O. Using the integral density method, we computed and thoroughly discussed the multifractal spectrum dimensions, studying the intricate distributional features at both partial and global scales over different nucleon density distributions within a nucleus. The multifractal dimension spectrum shows distinctive distribution characteristics for nucleon distributions with different structures. In the case of a relatively uniform nucleon distribution, as observed in stable nuclei, the spectrum shows a slight tendency to change, suggesting a mono-structure distribution. On the other hand, structures such as halo or cluster nuclei show pronounced changes in their spectral curves, indicating complex nucleon distribution structures. The analysis results underline the remarkable sensitivity of the multifractal dimension spectrum to the density distribution, highlighting its potential as a recommended feature variable for effectively describing the intricate characteristics of the nuclear density distribution.

\section*{Acknowledgements}
This work is supported by the Natural Science Foundation of Shanghai with Grants No.23JC1400200, the National Natural Science Foundation of China with Grants No.11905036,  12147101 and and 11890710, the National Key R\&D Program from the Ministry of Science and Technology of China (2022YFA1604900), and the STCSM under Grant No. 23590780100.

\end{CJK*}

\appendix*

\end{document}